\documentstyle[aps,pra,epsfig]{revtex}

\begin{document}

\title{Sum rules for hadronic elastic scattering for
the Tevatron, RHIC and LHC}

\author{Giulia Pancheri$^1$, Yogendra Srivastava$^{2,3}$ and Nicola Staffolani$^2$}

\address{$^1$ Laboratori Nazionali di INFN, Frascati, Italy }
\address{$^2$ Dipartimento di Fisica and INFN, Universit\`a di Perugia, Italy}
\address{$^3$ Physics Department, Northeastern University, Boston MASS USA}

\date{\today}

\maketitle

\begin{abstract}
Under the assumption of total absorption and dominance of the
imaginary part of the scattering amplitude, we present a sum rule for
any hadronic elastic differential cross-section ${{d\sigma}\over{dt}}$:
the dimensionless quantity ${{1}\over{2}}\int (dt)\sqrt{{{1}\over
{\pi}}{{d\sigma}\over{dt}}}\ \rightarrow\ 1$, at asymptotic energies.
Experimental data from ISR and Tevatron confirm a trend towards its
saturation and some estimates are presented for LHC. Its universality
and further consequences for the nature of absorption in QCD based models
for elastic and total cross-sections are explored. \\
\end{abstract}


Ab initio calculations of hadronic elastic amplitudes and total 
cross-sections (in QCD) are presently difficult due to our meager
understanding of ``soft'' physics, that is, the non-perturbative 
and confinement region of QCD. Hence, the need to invoke general 
principles such as analyticity and unitarity to obtain bounds
and restrictions on these amplitudes \cite{froissart,martin,cheung}.
Analyticity and unitarity 
are expected to hold for finite-ranged hadron dynamics, only 
massive hadrons being the bound states of quarks and glue. In the 
following, we find that, under rather mild assumptions, a universal 
behavior for all hadrons is likely to emerge at asymptotic energies.\\  

Consider the amplitude for an elastic process $A(p_a)\ +\ B(p_b)\ 
\rightarrow\ A(p_c)\ +\ B(p_d)$. Let $s\ =\ (p_a + p_b)^2$, be the 
square of the CM energy, $t\ =\ (p_a - p_c)^2$, be the momentum 
transfer and let us normalize the amplitude so that the differential 
and total cross-sections are given by
$$ 
   {{d\sigma}\over{dt}} = \pi |F(s,t)|^2;\ \ \ 
\sigma_{tot}(s) = 4\pi \Im m F(s, t=0). \eqno(1) \\
$$

To enforce (direct or $s$-channel) unitarity and incorporate the
knowledge that most of the hadronic scatterings at high energies
are peaked in the forward direction, an eikonal formalism is 
convenient. The elastic amplitude may be expanded
in the impact parameter ($b$- space) in the usual fashion
$$ 
F(s,t) = i\int_0^\infty (bdb)J_0(b\sqrt{-t})
\tilde{F}(s,b),\eqno(2a)
$$ 
in terms of the ``partial $b$-wave'' amplitudes
$$ 
\tilde{F}(s,b)= 1-\eta(s,b) e^{2i\delta_R(s,b)} ,\eqno(2b)
$$
where the inelasticity factor $\eta(s,b)$ lies between 0 and 1,
and $\delta_R(s,b)$ is the real part of the phase shift. The
dimensionless ``$b$-wave cross-sections'' are given by
$$ 
{{d^2\sigma_{el}}\over{d^2{\bf b}}} = 1-2\eta(s,b)\cos\{2\delta_R(s,b)\} +  
\eta^2(s,b) ,\eqno(3a)
$$
$$  
{{d^2\sigma_{inel}}\over{d^2{\bf b}}} = 1-\eta^2(s,b) ,\eqno(3b)
$$
and 
$$ 
{{d^2\sigma_{tot}}\over{d^2{\bf b}}} = 2[1- \eta(s,b)\cos\{2\delta_R(s,b)\}] .
\eqno(3c) \\
$$

Eqs.(3) show explicitly the maximum permissible rise for the different
cross-sections due to unitarity. For complete absorption of ``low''
partial waves at asymptotic energies (which translates into 
$\eta(s,b)\rightarrow\ 0$ for $b\rightarrow\ 0$ and $s\rightarrow\ 
\infty$), one obtains the geometric limit (including the contribution
from shadow scattering): 
$$
{{d^2\sigma_{el}}\over{d^2{\bf b}}} =
{{d^2\sigma_{inel}}\over{d^2{\bf b}}} = 
{{1}\over{2}} {{d^2\sigma_{tot}}\over{d^2{\bf b}}} \rightarrow\ 1
\ for\ b\rightarrow\ 0\ and\ s\rightarrow\ \infty. \eqno(4) \\
$$

Evidence for such a maximum rise (i.e., the validity of 
Eq.(4)) has been provided through various models, 
such as the resummed 
soft gluon models \cite{grau,corsetti,pacetti,godbole} and other 
models  \cite{block,gaisser,donnachie}, all of whom incorporate 
the observed rise in $pp$ and $p\bar{p}$ total cross-sections.\\ 

Our objective in the present paper is to provide 
model independent predictions through sum rules over experimentally 
measurable quantities such as ${{d\sigma}\over{dt}}$.
We shall first derive a lower bound  for the dimensionless
integral $I_0(s)$, defined as

$$ 
I_0(s) = {{1}\over{2}}\int_{-\infty}^0 (dt) \sqrt{{{d\sigma}\over{\pi dt}}}\;\;\;\;. \eqno(5) \\ 
$$

Using Eqs.(1) and (2), it is easy to show that

$$
I_0(s) \ge 1 - \eta(s,0). \eqno(6) \\
$$

To obtain an upper bound for $I_0(s)$, via the obvious inequality
$$
I_0(s) \le \int_0^\infty (qdq) \left[ |Im F(s, q^2)| + |Re F(s, q^2)| \right],
\eqno(7)
$$
further input are needed. Since $F(s, q^2)$ is analytic in $q^2\ =\ -t$
until $q^2\ =\ - \mu_0^2$ (the lowest mass exchanged in the $t$- channel),
the $b$- expansion must converge until this imaginary $q\ =\ i \mu_0$.
For positive values of $t$, in Eq.(2) the Bessel functions $J_0(b\sqrt{-t})$
become $I_0(b\sqrt{t})$. It is easily shown that the convergence of the 
$b$- expansion requires that \cite{pacetti} (for large values of $b$)
$$
\eta(s,b) cos\{2\delta_R(s,b) \} \rightarrow 1 - e^{- b^2\mu_1^2/2}, \eqno(8a)
$$
and
$$
\eta(s,b)|sin\{2\delta_R(s,b) \}| \rightarrow e^{- b^2 \mu_2^2/2}, \eqno(8b) 
$$ 
where $\mu_{1,2}$ are related to $\mu_0$. (Eqs.(8) are essential ingredients
in obtaining the Martin-Froissart bound). Now we make the crucial assumption
about the lack of oscillations (at least to leading order). That is, we 
assume $sin\{2\delta_R(s,b) \}$ does not change sign. Using Eqs.(7-8) and the
lack of change of sign assumption, we obtain
$$
I_0(s) \le 1 + 2 \left| \tan\{2\delta_R(s,0) \}\right|. \eqno(9) \\
$$

We can relate $|tan\{2\delta_R(s, 0)\}|$ to the asymptotic value of the
$\rho$ - parameter and obtain finally the upper bound
$$
I_0(s) \le 1 + {{K}\over{ln(s/s_0)}}\; , \eqno(10)
$$
where $K$ is a positive constant. Combining Eqs.(6) and (10), we have
the asymptotic bounds
$$
1 - \eta(s,0) \le I_0(s) \le 1 + {{K}\over{ln(s/s_0)}}\; . \eqno(11) \\
$$

The above bounds have been obtained incorporating (i) positivity, 
(ii) unitarity, (iii) correct behavior near $b\ =\ 0$ and (iv) the asymptotic 
behavior $b\ \rightarrow\ \infty$. The lack of oscillations implies that
the physics of the inelasticity factor $\eta$ and the real part of
the phase shift $\delta_R$ becomes very simple at high energies. $\eta$
starts out being small in the central region ($b$ near zero) and then
monotonically increases to its asymptotic value $1$ (for large $b$), whereas
the real part of the phase shift has a maximum in the central region
and dies out monotonically in the peripheral region. Under these premises
and provisos, the bounds only require values of these quantities in the
central region and any explicit dependences on the mass parameters
$\mu_{1,2}$ disappear. As we shall discuss later, it is due to unitarity
relating how fast an amplitude decreases with momentum transfer (the ``slope
parameter'') to the strength of the amplitude (``coupling constants'' or
the prefactors).
The above becomes an equality and in fact $I_0(s)$ equals $1$ 
if, as argued previously, at high energies, $\eta(s,0)$ approaches zero.
Our sum rule then reads
$$
I_0(s) = {{1}\over{2}}\int_{-\infty}^0 (dt) \sqrt{{{d\sigma}\over{\pi dt}}}
\ \rightarrow\ 1,\ for\ s\rightarrow\ \infty. \eqno(12) \\
$$ 

$I_0(s)$ should rise from its threshold value $2|a_0|k\ \rightarrow\ 0$, 
where $a_0$ is the S-wave scattering length (complex for $p\bar{p}$) 
and $k$ is the CM 3-momentum, to its asymptotic value $1$ as $s$ 
goes to infinity.  In Fig.(1), we show a plot of this integral for
available data \cite{abe,amos1800,bozzo,breakstone,amaldi,akerlof,ayres,asad} 
on $pp$ and $p\bar{p}$ elastic scattering for
high energies  \cite{bultmann}.
Highest energy data at $\sqrt{s}\ =\ 1.8\ TeV$ for $p\bar{p}$ 
from the Fermilab Tevatron \cite{abe}, give an encouraging value of $0.98\ \pm\ 0.03$  
demonstrating that indeed the integral is close to its asymptotic
value of $1$. We expect it to be even closer to $1$ at the LHC (our 
extrapolation gives the value $0.99\pm 0.03$ for LHC).

\vspace{1cm}

\begin{figure}
\begin{center}
\mbox{
\epsfxsize=12truecm \epsffile{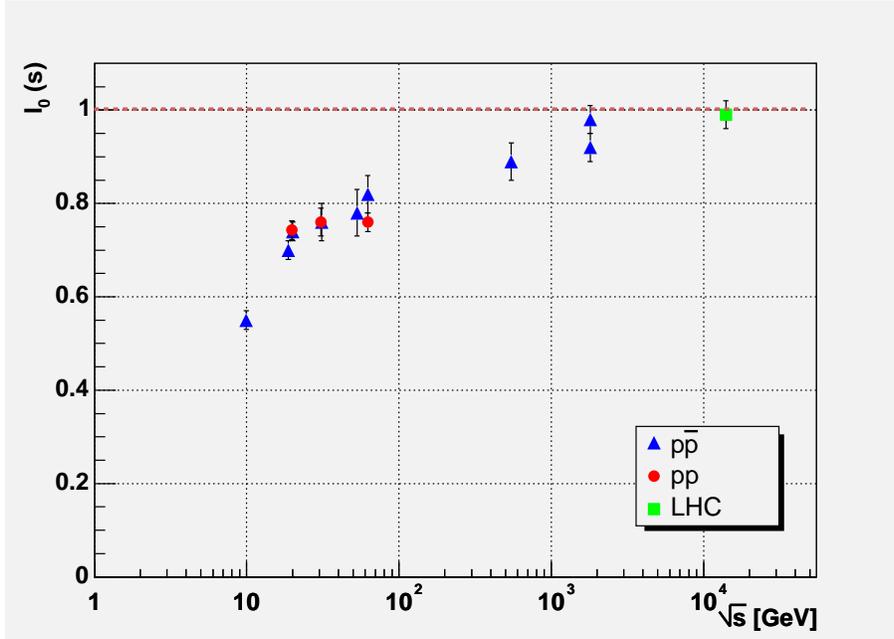}
}
\end{center}
\caption{\label{fig1}A plot of $I_0(s)$ vs. $\sqrt s$ using experimental data \protect\cite{abe,amos1800,bozzo,breakstone,amaldi,akerlof,ayres,asad}. 
The last point is our extrapolation for LHC.
 }
\end{figure}

Having found the trend to its asymptotic value at the highest 
available energies, we may return to ask whether the two assumptions 
made to obtain the sum rule are really necessary. Presently, we 
know theoretically that in the forward direction ($t=0$), 
the imaginary part must dominate the real part if the 
cross-section saturates the Froissart-Martin bound. That is, if at 
high energies \cite{khuri}  
$$
\sigma_{tot}(s) \rightarrow\ Constant \times ln^2(s/s_0), \ then\
\rho(s,0)\ =\ {{\Re e F(s,0)}\over{\Im m F(s,0)}}\ \rightarrow\ 
{{\pi}\over{ln(s/s_0)}}\ \rightarrow\ 0. \eqno(13) \\
$$

We can make a crude estimate of $\rho(s,0)$ through Eq.(13). For the
highest Tevatron energy, we find it to be about $0.2$. Since the
contribution of the real part to the integral only enters as $(1/2)
\rho^2$, the neglect of the real part in the forward direction would 
affect the integrand by about $2\%$, i.e., well within the 
experimental errors which are about $3\%$.\\
 
If the total cross-section were to increase only as $ln(s/s_0)$, the
ratio of real to the imaginary part would still go down to zero
as in Eq.(13), albeit with a smaller constant $\pi/2$. 
Thus, for  rising cross-sections,
we are assured of the correctness of our first assumption in the 
forward direction. For non-forward directions, we have no direct
evidence experimentally. However, since the overall differential 
cross-section would be decreasing (for $t\ \ne\ 0$) as a function
of $s$, their contribution to $I_0(s)$ is less important.
It is for exactly the same reason that the second assumption, i.e.
the absence of zeros in the imaginary part of the non-forward
amplitude, is not really necessary. So long as any possible such
zeroes remain at some finite values of (negative) $t$, they would 
not significantly upset our results. The satisfaction of our sum 
rule a posteriori justifies this claim.\\

There are some interesting and significant consequences which follow 
from the above analysis. First, since the asymptotic value is reached 
from below, we may bound $\eta$, the inelasticity  in the central 
($b\ =\ 0$) region. For example, even at $\sqrt{s}\ =\ 100\ GeV$,
absorption is not complete but only about $80\ \%$, giving us
a quantitative understanding of where the onset of ``high energy''
lies.\\

Another deduction concerning universality of the above result may be 
made. That is, the central value of inelasticity should approach 
zero for the scattering of all hadrons (at least for all
hadrons made of light quarks). Such a result follows 
naturally from QCD if we recall first the experimental fact that both for 
nucleons as well
as for mesons  \cite{ellis,barger}, half the hadronic energy 
is carried by glue.\\

In QCD, such an equipartition of energy has been derived rigorously 
to hold for all hadrons which are bound states of massless quarks \cite{widom}.
Since all available high energy elastic scattering data are for nucleons
and light mesons, all of which are made of the very light quarks, we have
excellent support from QCD for equipartition.
If one couples this with the notion that the rise of the 
cross-section is through gluon-gluon scattering, which is flavour 
independent, the asymptotic equality of (the rise in) all hadronic 
cross-sections 
automatically emerges.\\
 
Of course, the approach to asymptotia would not be the same between
nucleon-nucleon and meson-nucleon scatterings. It is unfortunate that 
data for $\pi N$ scattering are available only upto $\sqrt{s}\ =\ 20\ 
GeV$, which is far from asymptotic. In fact, for this channel, 
$I_0(s)$ is only about $0.6$ at the highest energy measured so far. 
Since the same asymptotic value of $1$ for this integral should be 
reached for all hadrons, the rise with energy must be even more 
dramatic for meson-nucleon scattering. In principle, such a test for 
RHIC and LHC may be feasible through Bjorken's suggestion \cite{bjorken} 
of converting an incident proton into a pion by isolating the one 
pion exchange contribution via tagging or triggering on a leading 
neutron or $\Delta^{++}$.\\

It appears reasonable to extend our analysis to $NA$ or even $AB$ 
elastic scatterings, where $A,B$ are nuclei. Also for these 
processes, at very high energies, we expect from QCD that the central 
inelasticity should approach zero and hence (modulo possible 
complications were $\rho(s,t)$ to be anomalously large), $I_0(s)$ 
should again asymptote to $1$.  For illustrative purposes, let us 
consider the following very simple expression which incorporates 
the sum rule:
$$
|F_{AB}(s,t)| = I_0(s){\cal B}_{AB}(s)e^{(1/2){\cal B}_{AB}(s) t}. 
\eqno(14) \\
$$

Parametrizations of the above form (which underestimate the large
$t$ contributions by ignoring the secondary slopes) are routinely 
used. However, what is new here is that since $I_0(s)$ goes to $1$ 
in the asymptotic limit, it is our prediction that the prefactor would, 
in the same limit, 
become equal to the diffraction width ${\cal B}_{AB}(s)$. 
Physically, it says that unitarity correlates and limits how large 
the amplitude can be, as a function of the energy, to how fast it 
decreases, as a function of the momentum transfer. If Eq.(14) holds, 
we may  use the optical theorem to obtain  the approximate expression
$$
I_0(s) = 
{{\sigma_{tot}(s)}\over{4\pi {\cal B}_{AB}(s)}}
\left[1+(1/2)\rho^2_{AB}(s,0)\right] . \eqno(15) \\
$$  

Under the same assumption, we would have 
$$
{{\sigma_{el}(s)}\over{\sigma_{tot}(s)}} \approx 
{{I_0(s)}\over{4}}
\left[1 + \rho^2_{AB}(s,0)\right]. \eqno(16) \\
$$

For the highest Tevatron energy $\sqrt{s}\ =\ 1.8\ TeV$, Eq.(16)
would estimate the elastic to total ratio to be about $0.25$ in
excellent accord with the experimental value $0.25\ \pm\ 0.02$  \cite{amos2}.\\

Future experiments from RHIC and LHC should be able to test our
sum rule predictions for $pp$, $p\bar{p}$ and hopefully other elastic 
channels. For this purpose, it would be useful if
experimentalists would present values of $I_0(s)$ directly from their 
experimental data, obviating thereby interpolations 
(such as those carried out by us to obtain FIG.~\ref{fig1}).\\

We acknowledge support from the European Union Contract number 
HPRN-CT2002-00311.


\begin{thebibliography}{99}

\bibitem{froissart}
M. Froissart, {\it Phys. Rev.} {\bf 123} (1961) 1053.

\bibitem{martin}
A. Martin, {\it Phys. Rev.} {\bf 129} (1963) 1432.

\bibitem{cheung}
A. Martin and F. Cheung, ``Analyticity properties and 
bounds on scattering amplitudes'', Gordon and Breach, Science
Publishers, New York (1970).

\bibitem{grau}
A. Grau, G. Pancheri and Y.N. Srivastava, {\it Phys. Rev.} {\bf D60} (1999) 114020; A. Corsetti, A. Grau, G. Pancheri and Y.N. Srivastava, 
{\it Phys. Lett.} {\bf B382} (1996) 282.

\bibitem{corsetti}
A. Corsetti, A. Grau, R.M. Godbole, G. Pancheri and Y.N. Srivastava, 
ArXiv: hep-ph/9605254; R.M. Godbole, A. Grau, G. Pancheri and 
Y.N. Srivastava, Invited talk at the International Workshop on QCD, 
Martina Franca, Italy, June, 2001. ArXiv: hep-ph/0205196.

\bibitem{pacetti}
A. Grau, S. Pacetti, G. Pancheri and Y.N. Srivastava,
{\it Nucl. Phys. B (Proc. Suppl.)} {\bf 126} (2004) 84.

\bibitem{godbole}
R.M. Godbole, A. Grau, G. Pancheri and Y.N. Srivastava,
{\it Nucl. Phys. B (Proc. Suppl.)} {\bf 126} (2004) 94.

\bibitem{block}
M.M. Block, E.M. Gregores, F. Halzen and G. Pancheri, 
{\it Phys. Rev.} {\bf D60} (1999) 054024; M.M. Block, {\it Nucl. Phys. 
B (Proc. Suppl.)} {\bf 126} (2004) 76; M.M. Block and K. Kang,
ArXiv: hep-ph/0302146.

\bibitem{gaisser} 
T. Gaisser and F. Halzen, {\it Phys. Rev. Lett.} {\bf 54} (1985) 1754. 

\bibitem{donnachie} A. Donnachie and P.V. Landshoff, {\it Phys. Lett.} {\bf B550} 
(2002) 160.


\bibitem{abe} 
F. Abe {\it et al.}, {\it Phys. Rev.} {\bf D50} (1994) 5518.

\bibitem{amos1800} 
N.A. Amos {\it et al.}, {\it Phys. Lett.} {\bf B247} (1990) 127; 
{\it Nucl. Phys.} {\bf B262} (1985) 689.  

\bibitem{bozzo} 
M. Bozzo {\it et al.}, {\it Phys. Lett.} {\bf B155} (1985) 197; 
{\it Phys. Lett.} {\bf B147} (1984) 392. 


\bibitem{breakstone}
A. Breakstone {\it et al.}, {\it Nucl. Phys.} {\bf B248} (1984) 253;
{\it Phys. Rev. Lett.} {\bf 54} (1985) 2160.  

\bibitem{amaldi}
U. Amaldi and K.R. Schubert, {\it Nucl. Phys.} {\bf B166} (1980) 013. 

\bibitem{akerlof}
C. Akerlof {\it et al.}, {\it Phys. Rev.} {\bf D14} (1976) 2864.

\bibitem{ayres}
D. Ayres {\it et al.}, {\it Phys. Rev.} {\bf D15} (1977) 3105.

\bibitem{asad}
Z. Asad  {\it et al.}, {\it Phys. Lett.} {\bf B108} (1982) 51.  

\bibitem{bultmann}
Interesting new data on $pp$ elastic scattering at $\sqrt{s}\ =\
200\ GeV$ have recently been published (S. B\"ultmann  {\it et al.},
ArXiv: nucl-ex/0305012). However, these data cover 
only the very forward region ($|t|\ <\ 0.019\ GeV^2$) and hence are moot 
regarding our sum rule.
\par\noindent

\bibitem{khuri}
N.N. Khuri and T. Kinoshita, {\it Phys. Rev.} {\bf B137} (1965) 720.

\bibitem{ellis}
W.K. Ellis, W.J. Stirling and W.R. Webber, ``QCD and Collider Physics", 
Cambridge University Press, New York (1996).

\bibitem{barger}
V. Barger and R.J. Phillips, ``Collider Physics", Addison Wesley 
Publishing Co., California (1987). 

\bibitem{widom}
Y.N. Srivastava and A. Widom, {\it Phys. Rev.} {\bf D63} (2001) 077502. 

\bibitem{bjorken}
J.D. Bjorken, {\it Nucl. Phys. B (Proc. Suppl.)} {\bf 71} (1999) 484. 

\bibitem{amos2}
N.A. Amos {\it et al.}, {\it Phys. Rev. Lett.} {\bf 63} (1989) 2784.

\end{thebibliography}
\end{document}